\documentclass[aps,prl,twocolumn,showpacs, superscriptaddress, groupedaddress,nofootinbib]{revtex4}  
\usepackage{graphicx}  
\usepackage{dcolumn}   
\usepackage{bm}        
\usepackage{amssymb, amsmath}
\usepackage{slashed}   

\usepackage[usenames,dvipsnames,svgnames,table]{xcolor}

\newcommand{\GeV}{{\rm GeV}}

\newcommand{\eps}{\epsilon}
\def\beq{\begin{equation}}
\def\eeq{\end{equation}}

\begin{document}
\widetext


\hspace{5.2in} \mbox{TTP16-058, CERN-TH-2016-261}

\title{The Axiflavon}
\author{Lorenzo Calibbi}
\affiliation{CAS Key Laboratory of Theoretical Physics and Kavli Institute for Theoretical Physics
China (KITPC), Institute of Theoretical Physics, Chinese Academy of Sciences,
Beijing 100190, P. R. China}
\author{Florian Goertz}
\affiliation{Theory Division, CERN, CH-1211 Geneva 23, Switzerland}
\author{Diego Redigolo}
\affiliation{Raymond and Beverly Sackler School of Physics and Astronomy, Tel-Aviv University, Tel-Aviv
69978, Israel}
\affiliation{Department of Particle Physics and Astrophysics, Weizmann Institute of Science, Rehovot 7610001,Israel}
\author{Robert Ziegler}
\affiliation{Institute for Theoretical Particle Physics (TTP), Karlsruhe Institute of Technology,
Engesserstrasse 7, D-76128 Karlsruhe, Germany}
\author{Jure Zupan}
\affiliation{Theory Division, CERN, CH-1211 Geneva 23, Switzerland}
\affiliation{Department of Physics, University of Cincinnati, Cincinnati, Ohio 45221,USA}

\begin{abstract}
\noindent
We show that solving the flavor problem of the Standard Model with a simple $U(1)_H$ flavor symmetry 
naturally leads to an axion that solves the strong CP problem and constitutes a viable Dark Matter candidate. In this framework, the ratio of the axion mass and its coupling to photons is related to the SM fermion masses and predicted within a small range, as a direct result of the observed hierarchies in quark and charged lepton masses. The same hierarchies determine  the axion couplings to fermions, making the framework very predictive and experimentally testable by future axion and precision flavor experiments.

\end{abstract}

\maketitle
\noindent
Three of the major open questions in particle physics are (i) the strong CP problem -- why is the QCD $\theta$ angle so small, 
(ii) what is the origin of Dark Matter (DM),  and (iii) the Standard Model (SM) flavor puzzle -- why are the masses of fermions so hierarchical. 
The first problem can be elegantly addressed by the QCD axion: the pseudo Goldstone boson of an approximate global $U(1)$ symmetry that has a color anomaly \cite{Peccei:1977hh,Wilczek:1977pj,Weinberg:1977ma}. The two main classes of axion models based on this mechanism are usually referred to as the KSVZ~\cite{Kim:1979if, Shifman:1979if} and the DFSZ~\cite{Dine:1981rt, Zhitnitsky:1980tq} axion solutions. It is well known~\cite{Preskill:1982cy, Abbott:1982af, Dine:1982ah} that in most regions of the parameter space the QCD axion serves as a viable DM candidate. Finally, the SM flavor problem can be elegantly resolved  by introducing approximate flavor symmetries, which are spontaneously broken at large scales as in the original Froggatt-Nielsen (FN) mechanism~\cite{Froggatt:1978nt}.

In this letter we propose a unified framework where the approximate symmetry of the QCD axion is identified with the simplest flavor symmetry of the FN mechanism (the setup is the minimal realization of an old idea by F.~Wilczek~\cite{Wilczek:1982rv} that axion and flavor physics could be connected). The structure of quark and lepton masses and mixings follows from a spontaneously broken $U(1)_H$ flavor symmetry which generically has a QCD anomaly. The resulting Nambu-Goldstone boson,  the {\it axiflavon}, solves automatically the strong CP problem by dynamically driving the theory to a CP conserving minimum \cite{Vafa:1984xg}. Non-thermal production of the axiflavon from the misalignment mechanism can then reproduce the observed DM relic density, provided that the $U(1)_H$ breaking scale is sufficiently large.

This simultaneous solution of flavor, strong CP and DM problem leads to sharp predictions for the properties of the axiflavon that can be tested experimentally.\footnote{A similar approach has been proposed in Ref.~\cite{Giudice:2012zp}, where the requirement of gauge coupling unification was combined with the KSVZ axion solution to strong CP and DM problem in order to determine the phenomenology of the so-called unificaxion. }
Of particular importance is the axion coupling to photons that is determined by the ratio $E/N$, i.e., the ratio of the electromagnetic over the QCD anomaly coefficient. This ratio is essentially a free parameter in generic axion models (see Refs.~\cite{DiLuzio:2016sbl,Farina:2016tgd} for a recent discussion). In the axiflavon setup $E/N$ is directly related to the $U(1)_H$ charges of SM fermions and thus to the hierarchies between SM fermion masses. Despite the considerable freedom of choosing these charges in the simplest $U(1)_H$ model, we find a surprisingly sharp prediction for $E/N$ centered around 8/3, the prediction of the simplest DFSZ model, 
\begin{align}
\frac{E}{N} \in [2.4 , 3.0] \, .
\end{align} 
This result is a direct consequence of the strong hierarchies in up- and down-type quark masses and only weak hierarchies in the ratio of down-quark to charged lepton masses. A similarly restrictive range for $E/N$ can be found also in a broad class of models with non-minimal flavor symmetries like $U(2)$ (which are more predictive in the fermion sector).  The above range for $E/N$ can be translated into a prediction for the ratio of axion-photon coupling $g_{a \gamma \gamma}$ and axion mass $m_a$
\begin{align}
\frac{g_{a \gamma \gamma}}{m_a} \in  \frac{\left[ 1.0, 2.2 \right]}{10 ^{16} \GeV}  \frac{1}{ \mu {\rm eV}} \, .
\end{align}
For axion masses in the natural range for axion DM, $m_a \approx (10^{-3} \div 0.1) \,{\rm meV}$, this region will be tested in the near future by the ADMX experiment.  

The axiflavon can also be tested by precision flavor experiments looking for the decay $K^+ \to \pi^+ a$. Indeed the flavor violating couplings of the axiflavon to quarks are also related to quark masses, but in contrast to $E/N$ are more sensitive to model-dependent ${\cal O}(1)$ coefficients
\begin{equation}
{\rm BR} (K^+ \to \pi^+ a)\simeq 1.2 \cdot 10^{-10} \left( \frac{m_a}{0.1 \, {\rm meV} } \right)^2 \left( \frac{\kappa_{sd}}{N} \right)^2   \, , \end{equation}
where $\kappa_{sd}/ N \sim {\mathcal O}(1)$. In the natural range of axion DM this decay can be within the reach of the NA62 and ORKA experiments, depending on the model-dependent coefficients. We summarize our results along with the present and expected experimental constraints in Fig.~1 at the end of this letter.

\section{Setup} We assume that the masses of the SM fermions come from the vacuum expectation value (vev) $v=174$\, GeV of the SM Higgs $H$, while the hierarchies of the Yukawa couplings are due to a global horizontal symmetry $U(1)_H$.  The SM Weyl fermion fields $Q_i, U_i^c, D_i^c, L_i, E_i^c$ have positive flavor-dependent charges $[q]_i, [u]_i, [d]_i, [l]_i, [e]_i$, respectively. Here $Q_i$ and $L_i$ are the quark and lepton electroweak doublets, the remaining fields are $SU(2)_L$ singlets, and $i=1,2,3$ is the generation index. For simplicity we assume that the Higgs does not carry a $U(1)_H$ charge, so that the flavor hierarchies are explained entirely by the fermion sector. This assumption will be relaxed below. The $U(1)_H$ symmetry is spontaneously broken at a very high scale by the vev $V_\Phi$ of a complex scalar field $\Phi$ with $U(1)_H$ charge of $-1$. All other fields in the model, the FN messengers, have masses of ${\mathcal O}(\Lambda) \gtrsim V_\Phi \gg v$ and can be integrated out. Note that $\Lambda$ is  a scale above $U(1)_H$ breaking, implying that fermionic FN messengers are vector-like under the $U(1)_H$. The Yukawa sector in the resulting effective theory is then given by 
\begin{align}
{\cal L} &  = a^u_{ij} Q_i U^c_j H \left( \Phi/ \Lambda\right)^{[q]_i + [u]_j } +  a^d_{ij} Q_i D^c_j \tilde{H} \left( \Phi/\Lambda \right)^{[q]_i + [d]_j } \nonumber \\
&+  a^e_{ij} L_i E^c_j \tilde{H} \left( \Phi/\Lambda \right)^{[l]_i + [e]_j }  + {\rm h.c.} \, ,
\label{eq:lagrangian}
\end{align}
where $a^{u,d,e}_{ij}$ are complex numbers, assumed to be ${\cal O}(1)$. Setting $\Phi$ to its vev, $\langle \Phi\rangle=V_\Phi/\sqrt 2$, gives the SM Yukawa couplings with
\begin{align}
y^{u,d,e}_{ij} & = a^{u,d,e}_{ij} \eps^{[L]_i + [R]_j} \,,
\end{align}
where $[L]_i = [q]_i, [R]_i = [u]_i, [d]_i$ in the quark sectors, $[L]_i = [l]_i, [R]_i = [e]_i$ in the charged lepton sector and we have defined the small parameter $\eps \equiv V_\Phi/(\sqrt 2 \Lambda)$. 

The hierarchy of masses follows from $U(1)_H$ charge assignments, giving $y_{ij}^f\sim \hat{V}_{ij} m_j^f/v$, with $m_i^f$ the SM fermion masses and $\hat{V}_{ij} = V_{ij}$ for $ i \le j$, $\hat{V}_{ij} = 1/V_{ij}$ for $ i \ge j$. Here $V$ is the CKM matrix in the quark sector and the PMNS matrix in the charged lepton sector.  
 The observed CKM structure is typically obtained for $\epsilon$ of the order of the Cabibbo angle, $\eps \sim 0.23$. The exact values of $U(1)_H$ charges can be obtained from a fit to fermion masses and mixings, and are subject to the uncertainties in  the unknown ${\cal O}(1)$ numbers $a^{u,d,e}_{ij}$. As we are going to demonstrate, these uncertainties will only weakly influence the main phenomenological predictions. Note that the pattern of masses and mixings in the neutrino sector can also be explained in this setup, however, this sector of the SM is irrelevant for the prediction of color and electromagnetic $U(1)_H$ anomalies.

The field $\Phi$ contains two excitations, the CP-even flavon, $\phi$, and the CP-odd axiflavon, $a$,
\beq
\Phi=  \frac{1}{\sqrt 2}(V_\Phi + \phi\big) e^{ia/V_\Phi}.
\eeq
The flavon field $\phi$ has a mass $m_\phi\sim{\cal O}(V_\Phi)$, and  thus is not directly relevant for low energy phenomenology, and can be integrated out. The axiflavon, $a$, is a Nambu-Goldstone boson. It is  massless at the classical level, but receives a nonzero mass from the breaking of $U(1)_H$ by the QCD anomaly.
Its couplings to SM fermions $F_i$ are given by
\begin{align}
{\cal L}_{a f f} & = \lambda_{ij}^f  a  F_i F^c_j + {\rm h.c.} \, ,  
\end{align}
with
\begin{gather}
\lambda_{ij}^{u,d,e}  = i ([L]_i + [R]_j)\frac{v}{ V_\Phi } y^{u,d,e}_{ij} \, . \label{eq:lambdas}
\end{gather}
The couplings of the axiflavon to the SM fermions are in general not diagonal in the fermion mass eigenstate basis 
due to the generation-dependency of charges $[q]_i, [u]_i, [d]_i, [l]_i, [e]_i$. This induces flavor changing neutral currents (FCNCs), which are experimentally well constrained and will be discussed in the next section\footnote{For flavor constraints on a heavy CP-odd flavon 
and possible collider signatures see Ref.~\cite{Bauer:2016rxs}.}. Note that several axion models with flavor-violating couplings to fermions have been proposed in the literature, see e.g.~\cite{Reiss:1982sq, Davidson:1984ik, Berezhiani:1990wn, Berezhiani:1990jj, Babu:1992cu, Feng:1997tn, Albrecht:2010xh, Ahn:2014gva, Celis:2014iua}.  In the axiflavon setup they are directly related to the SM fermion masses and thus predicted up to ${\cal O}(1)$ uncertainties. 

The axiflavon  couplings to gluons and photons are controlled by the color and electromagnetic anomalies,
\begin{align}
{\cal L} & =  \frac{\alpha_s}{8 \pi} \frac{a}{f_a} G \tilde{G} +  \frac{E}{N} \frac{\alpha_{\rm em}}{8 \pi} \frac{a}{f_a} F \tilde{F} \, , 
\end{align}
where $\tilde{G}_{\mu \nu} = \frac{1}{2} \eps_{\mu \nu \rho \sigma} G^{\rho \sigma}$ and we have switched to the standard axion notation introducing $f_a = V_\Phi/2 N$. The two anomaly coefficients, $N,E$, are completely determined by the $U(1)_H$ charges of SM fermions
\begin{gather}
N  = \frac{1}{2} \sum_i 2 [q]_i + [u]_i + [d]_i \, , \\ 
\!\! E  = \sum_i  \frac{4}{3} \left( [q]_i + [u]_i \right) + \frac{1}{3} \left( [q]_i + [d]_i \right) + [l]_i + [e]_i \, , 
\end{gather} 
in the minimal scenario where these are the only states with chiral $U(1)_H$ charge assignments (see a more detailed discussion below). 
Interestingly, these coefficients can be directly related to the determinants of the fermion mass matrices as \cite{Ibanez:1994ig, Binetruy:1994ru, Binetruy:1996xk}
\begin{align}
\label{detmumd}
{\rm det} \,  m_u \, {\rm det} \,  m_d & = \alpha_{ud} \, v^6 \eps^{2 N} \, , \\
 {\rm det} \,  m_d / {\rm det} \,  m_e & =  \alpha_{de} \, \eps^{\frac{8}{3} N - E}  \, , 
\end{align}  
where the quantities $\alpha_{ud} = {\rm det} \,  a_u {\rm det} \, a_d $ and $\alpha_{de} = {\rm det} \,  a_d/{\rm det} \,  a_e$ contain the ${\cal O}(1)$ uncertainties, given by the anarchical coefficients in Eq.~(\ref{eq:lagrangian}). Taking fermion masses at $10^9 \, \GeV$ from Ref.~\cite{Xing:2007fb}, one finds ${\rm det} \,  m_u {\rm det} \,  m_d/v^6 \approx 5 \cdot 10^{-20} $ and  ${\rm det} \,  m_d/ {\rm det} \, m_e \approx 0.7$, which makes it clear that up to small model-dependent corrections we have $E = 8/3\, N$ and so are close to the simplest DFSZ axion solution~\cite{Kim:1998va}. Indeed the phenomenologically relevant ratio $E/N$ is independent of $\eps$ and given by
\begin{align}
\frac{E}{N} & = \frac{8}{3} - 2 \frac{\log \frac{{\rm det} \,  m_d}{{\rm det} \,  m_e} - \log \alpha_{de}}{\log \frac{{\rm det} \,  m_u {\rm det} \,  m_d}{v^6} - \log \alpha_{ud}} \, .\label{ENprediction}
\end{align}
 The most natural values for the coefficients are $\alpha_{ud}  = \alpha_{de} = 1$, in the sense that Yukawa hierarchies are entirely explained by $U(1)_H$ charges, giving $E/N \approx 2.7$. To estimate the freedom from ${\cal O}(1)$ uncertainties, we simply take flatly distributed ${\cal O}(1)$  coefficients in the range $[1/3,3]$ with random sign, resulting in a 99.9\% range 
\begin{align}
\frac{E}{N} & \in \left[ 2.4 , 3.0 \right] \, ,  
\label{ENpred}
\end{align}
or $| \frac{E}{N} -1.92|\in \left[ 0.5 , 1.1 \right]$,  to be compared with the usual KSVZ/DFSZ axion window $|\frac{E}{N} -1.92|  \in \left[ 0.07 , 7 \right] $~\cite{Nakamura:2010zzi}. 
Note that the restricted range is due to the suppression of the second term in Eq. \eqref{ENprediction} since the denominator is dominated by $\log {\rm det} \,  m_u {\rm det} \,  m_d/v^6 \approx -44 $, while the first term in the numerator is $\log  {\rm det} \,  m_d/ {\rm det}\, m_e \approx -0.36$. Following Ref.~\cite{diCortona:2015ldu}, we therefore obtain a quite sharp prediction for the axion-photon coupling,  $\frac{1}{4} g_{a \gamma \gamma}a F \tilde{F}$,  as
\begin{align}
g_{a \gamma \gamma} \in  \frac{\left[ 1.0, 2.2 \right]}{10 ^{16} \GeV}  \frac{m_a}{\mu {\rm eV}} \, ,\label{photonocoupling}
\end{align}
while the axion mass induced by the QCD anomaly is given by~\cite{diCortona:2015ldu}
\begin{align}
m_a & = 5.7 \, {\rm \mu eV} \left( \frac{10^{12} \GeV}{f_a} \right) \, .
\end{align}
It is remarkable that the prediction for $E/N$ in Eq.~(\ref{ENpred}) is largely insensitive on the details of the underlying flavor model. We therefore briefly review the underlying assumptions that lead to the above results and discuss their relevance and generality. First of all we are assuming positive fermion charges. This assumption can be relaxed to the extent that just the sums of charges in each Yukawa entry are positive, or equivalently that only $\Phi$ enters in the effective operators but not $\Phi^*$. This assumption follows naturally from holomorphy of the superpotential, if we embed the setup into a  supersymmetric model in order to address also the hierarchy problem. Our second assumption was that only the fermion fields and the flavon carry the $U(1)_H$ charges.
This assumption can be easily dropped since a possible $U(1)_H$ charge for the Higgs, $[h]$, would simply drop out of Eq.~(\ref{ENpred}), as it would enter as ${\rm det} \, m_u \to {\rm det} \, m_u \eps^{3[h]}$ and ${\rm det} \, m_{d,e} \to {\det} \, m_{d,e} \eps^{-3[h]}$.  Finally we have assumed that only light fermions contribute to the QCD and electromagnetic anomalies, i.e., that all the other fields in the model are either bosons or vectorlike fermions under $U(1)_H$. This is a natural feature of the FN messengers needed to UV-complete the effective setup in Eq.~(\ref{eq:lagrangian}), see also the explicit UV completions in Refs.~\cite{SN1, Calibbi:2012yj}.

We also note that the same prediction for $E/N$ holds in any flavor model where a global, anomalous $U(1)$ factor determines exclusively the determinant of the SM Yukawa matrices. For example  in $U(2)$ flavor models~\cite{Barbieri:1995uv, Barbieri:1997tu,  Dudas:2013pja, Falkowski:2015zwa}, where the three fermion generations transform as ${\bf 2} + {\bf 1}$, one has a $SU(2)$ breaking flavon and a $U(1)$ breaking flavon. In the supersymmetric realization, or upon imposing positive charge sums in the non-SUSY realizations, one finds texture zeros for the 11, 13 and 31 entries of the Yukawa matrices. The determinant is therefore given by the 12, 21 and 33 entries which are $SU(2)$ singlets and therefore depend only on $U(1)$ charges, resulting in the same prediction for  $E/N$ when the $U(1)$ breaking flavon contains the axiflavon (and the $SU(2)$ is gauged). 

Finally we comment on the modification for the $E/N$ range in the context of an additional light Higgs doublet, restricting for simplicity to the case of a 2HDM of Type-II. Then Eq.~(\ref{detmumd}) is modified by the rescaling $v^6 \to \sin_\beta^3 \cos_\beta^3 v^6$ where $\tan_\beta = v_u/v_d$ is the ratio of Higgs vevs. Large values of $\tan_\beta$ can reduce the suppression of the model-dependent term in Eq.~(\ref{ENprediction}), and we find essentially the same 99.9\% ranges for $\tan_\beta = 20$, while for $\tan_\beta = 50$ the range is slightly increased,  $E/N \in [2.3,3.0]$.

\begin{figure*}[t]
\includegraphics[scale=0.6]{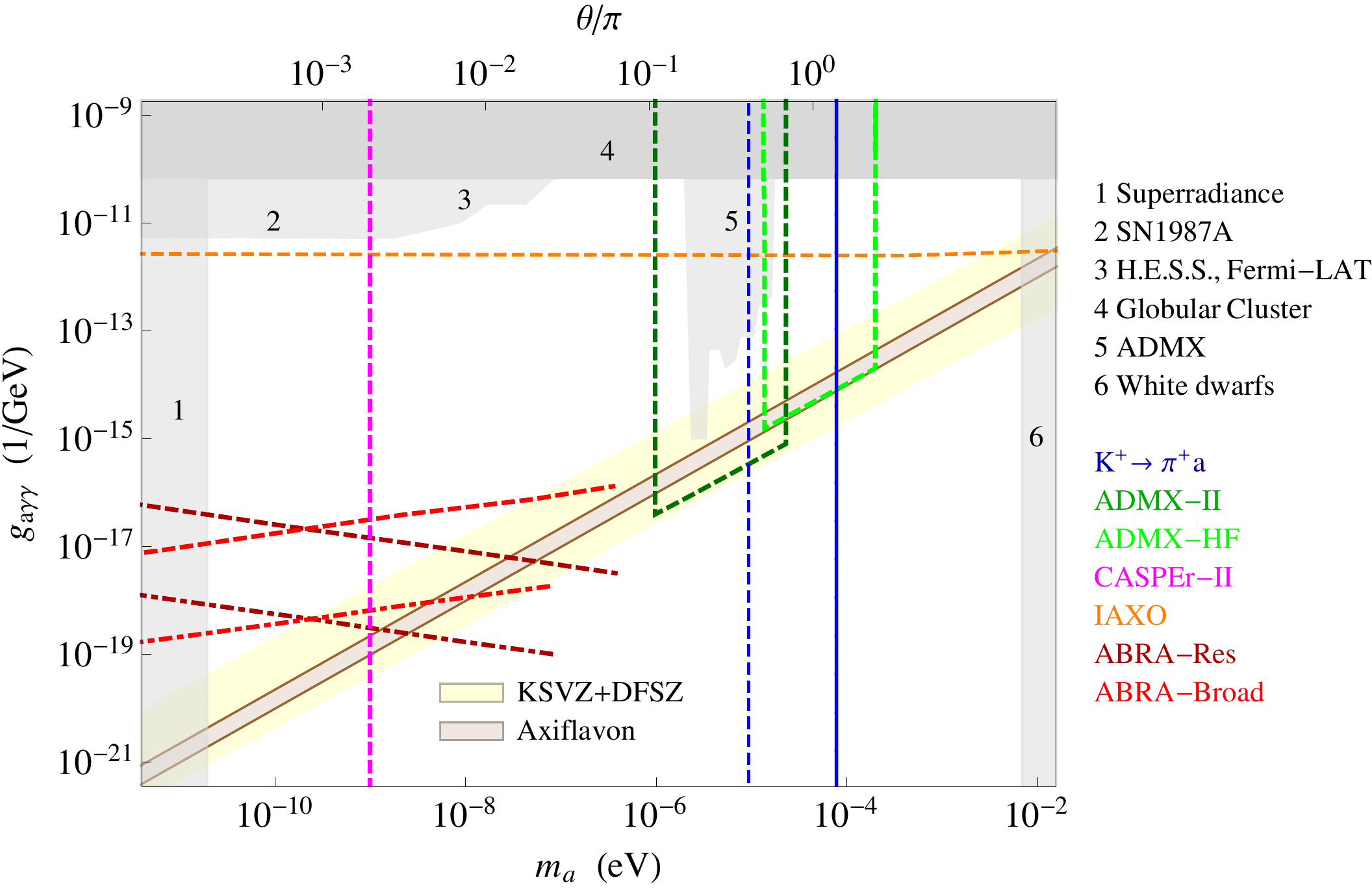}
\caption{The axiflavon band (light brown) projected on the axion parameter space: mass vs. photon coupling defined in Eq.~\eqref{photonocoupling}. The standard KSVZ/DFSZ band is shown in light yellow. The grey exclusion region is obtained from the combination of various axion constraints that are summarized in the legend. The dashed colored lines show the projected reach of future axion experiments. The solid blue line is the exclusion reach from current flavor experiments for an axiflavon model with $\kappa_{sd}/N = 1$ (cf.~Eq. \eqref{currentKbound}). The dashed blue line depicts the expected reach of future flavor experiments for the same choice of parameters. \label{fig:summary} 
}
\end{figure*}

\section{Phenomenology}
Being a QCD axion, the axiflavon is a very light particle with a large decay constant making it stable on cosmological scales. Assuming that the phase transition corresponding to the $U(1)_H$ breaking happens before inflation, the energy density stored in the axion oscillations can be easily related to the present Dark Matter (DM) abundance~\cite{Preskill:1982cy,Abbott:1982af,Dine:1982ah}: 
\begin{equation}
\Omega_{\rm DM}h^2\approx 1 \times10^{-7}\left(\frac{\text{eV}}{m_a}\right)^{7/6} \theta^2\ .
\end{equation}
For a given axion mass below roughly $\lesssim 10^{-5}-10^{-4}\text{ eV}$ it is then always possible to choose a misalignment angle $\theta$ to get the correct dark matter abundance $\Omega_{\rm DM}h^2\approx 0.12$. The axion domain wall problem is automatically solved in this setup, but interesting constraints can arise from isocurvature perturbations \cite{DEramo:2014urw}. 

We show in Fig.~\ref{fig:summary} present and future bounds on the axiflavon both from axion searches and from flavor experiments in terms of its mass  $m_a$ and its coupling to photons $g_{a\gamma\gamma}$. In this plane one can appreciate how the allowed range of $E/N$ is considerably reduced compared to the standard axion window \cite{Nakamura:2010zzi}. Assuming that the axiflavon is also accounting for the total DM abundance we give the corresponding value of $\theta$ for a given mass. 

In the high mass region with $m_a\sim 0.1-10\text{ meV}$ stringent bounds on the axiflavon comes from its coupling to fermions and are hence independent of $g_{a\gamma\gamma}$. A mild lower bound on the axiflavon decay constant $f_a$ can be derived from axiflavon coupling to electrons which affects white dwarf cooling \cite{Raffelt:1985nj}. This bound cuts off our parameter space at around $m_a\lesssim 10\text{ meV}$.

A stronger bound comes from the flavor-violating coupling of the axiflavon 
to down and strange quarks, $a \overline{s} d $, 
leading to (bounds from kaon decays are more restrictive than the bounds from kaon mixing)
\begin{equation}
\Gamma(K^+ \to \pi^+ a)\simeq \frac{m_K }{64 \pi} |\lambda_{21}^d +\lambda_{12}^{d*}|^2 B_s^2 \biggr(1-\frac{m_\pi^2}{m_K^2}\biggr),
\end{equation}
where $m_{K,\pi}$ are the kaon and pion masses, and $B_s=4.6(8)$ is the nonperturbative parameter related to the quark condenstate~\cite{Kamenik:2011vy}. The 90\% CL  combined bound from E787 and E949, ${\rm BR} (K^+ \to \pi^+ a)  < 7.3 \cdot 10^{-11} $ \cite{Adler:2008zza}, gives   
\beq
\frac{1}{2}|\lambda_{21}^d +\lambda_{12}^{d*}| < 1.4 \cdot 10^{-13}.
\eeq
Defining $|\lambda_{21}^d +\lambda_{12}^{d*}| \equiv 2 \kappa_{sd}\sqrt{m_d m_s}/(2 N f_a )$, this gives
\begin{align}
f_a \gtrsim \frac{\kappa_{sd}}{N}  \times 7.5 \cdot 10^{10} \, \GeV \, ,\label{currentKbound}
\end{align}
where $\kappa_{sd}/N \sim {\mathcal O}(1)$ 
are model-dependent coefficients controlled by the particular flavor charge assignments, and quark masses are taken at $\mu \sim 2 \, {\rm GeV}$. Similarly in the $B$ sector we find
\begin{align}
\Gamma(B^+ \to K^+ a)\simeq \frac{m_B}{64 \pi} |\lambda_{32}^d +\lambda_{23}^{d*}|^2  \big(f_0^K(0)\big)^2  \delta_{BK} \, ,
\end{align}
with $f_0^K(0)=0.331$~\cite{Ball:2004ye} and the shorthand notation
\beq
\delta_{BK}=\biggr(\frac{m_B}{m_b-m_s}\biggr)^2\biggr(1-\frac{m_K^2}{m_B^2}\biggr)^3.
\eeq
Defining $|\lambda_{32}^d +\lambda_{23}^{d*}| \equiv 2 \kappa_{bs}\sqrt{m_b m_s}/(2 N f_a )$, this gives for the branching ratio
\begin{equation}
{\rm BR} (B^+ \to K^+ a)\simeq 1.4 \cdot 10^{-12} \left( \frac{m_a}{0.1 \, {\rm meV} } \times  \frac{\kappa_{bs}}{N} \right)^2 \, , 
\end{equation}
where again $\kappa_{bs}/N \sim {\mathcal O}(1)$. A bound ${\rm BR} (B^+ \to K^+ a) < 10^{-6} \div 10^{-8}$, potentially in the reach of Belle II, would translate into $m_a < \left( 8 \div 80 \right) \, {\rm meV} \times N/\kappa_{bs}$. A careful experimental analysis of this decay would be very interesting, as suggested also in Ref.~\cite{Feng:1997tn}.

The solid blue line in Fig.~\ref{fig:summary} shows the lower bound on $m_a$ from flavor-violating kaon decays for $\kappa_{sd}/N=1$. The reach on ${\rm BR} (K^+ \to \pi^+ a)$ is expected to be improved by a factor $\sim 70$ by NA62~\cite{Anelli:2005ju, Fantechi:2014hqa} (and possibly also ORKA~\cite{Comfort:2011zz} and KOTO~\cite{Tung:2016xtx}), giving sensitivity to scales as high as $f_a \gtrsim\kappa_{sd}/N  \times 6.3 \cdot 10^{11} \, \GeV$. The expected sensitivity on the axion mass for $\kappa_{sd}/N=1$ is shown by the dashed blue line in Fig.~\ref{fig:summary}. Therefore future flavor experiment will probe the axiflavon parameter space in the interesting region where it can account for the dark matter relic abundance with $\theta\sim \mathcal{O}(1)$.

 Going to lower axiflavon masses, below $0.1\text{ keV}$, the phenomenology becomes essentially identical to the one of the original DFSZ model but with a sharper prediction for the value of $E/N$, given in Eq. (\ref{ENpred}). This corresponds to the brown band in Fig.~\ref{fig:summary}. 
 
 The gray shaded regions in Fig.~\ref{fig:summary} summarize the present constraints on axion-like particles. An upper bound on the photon coupling for the full range of masses of our interest comes from its indirect effects on stellar evolution in Globular Clusters \cite{Ayala:2014pea}. A comparable bound is set by the CAST experiment \cite{Andriamonje:2007ew}. Stronger constraints for axions lighter than $0.1 \text{ }\mu\text{eV}$ can be derived from the lack of a gamma-ray signals emitted from the supernova SN1987A \cite{Payez:2014xsa} and from the bounds on spectral irregularities of the Fermi-LAT and H.E.S.S. telescopes \cite{TheFermi-LAT:2016zue,Abramowski:2013oea}. The region of very low axion masses below $10^{-5}\text{}\mu\text{eV}$ is disfavoured by black hole superradiance independently on the photon coupling \cite{Arvanitaki:2014wva}. In the axion mass region between  $1 \text{ }\mu\text{eV}$  and $100\text{ }\mu\text{eV}$ present bounds from the ADMX experiment \cite{Asztalos:2009yp} do not put yet a constraint on the axiflavon band. This is a well-known feature of the original DFSZ model with $E/N=8/3$ that is shared by the axiflavon and further motivates future developments in microcavity experiments. 

In Fig. \ref{fig:summary} we also display the projections for the different axion future experiments. The combination of the upgraded ADMX experiment and its High Frequency version \cite{Shokair:2014rna} can probe a wide range of the axiflavon parameter space in the mass window between $1 \text{ }\mu\text{eV}$  and $100 \text{ }\mu\text{eV}$. This region is strongly preferred because the correct axion abundance can be obtained without a tuning of the initial misalignment angle. Dielectric Haloscopes \cite{TheMADMAXWorkingGroup:2016hpc} have a similar reach of ADMX-HF and are not displayed in the plot. The IAXO experiment \cite{Vogel:2013bta} gives instead a bound only at large axiflavon masses $m_a\gtrsim \text{meV}$.  Such large masses are already robustly ruled out by flavor-violating kaon decays. The low mass window of the axiflavon band for $m_a\lesssim 0.1\text{ }\mu\text{eV}$ will be probed by the resonant ABRACADABRA experiment and its upgrade \cite{Kahn:2016aff}. Interestingly, the axiflavon band lives below the reach of the first phase of the broadband ABRACADABRA experiment. Axiflavon masses below  $10^{-3}\text{  }\mu\text{eV}$ will eventually be probed in the final phase of the CASPEr experiment \cite{Budker:2013hfa}.  

In conclusion, the axiflavon parameter space is considerably narrower than that of KSVZ/DFSZ models, as visible in Fig.~\ref{fig:summary}, and will be covered in a wide range of masses  by a combination of future axion searches and kaon experiments. In the high mass window with $10^{-6}\text{eV}\lesssim m_a\lesssim 10^{-4} \text{ eV}$ the comparable projected reaches of ADMX-HF and future kaon experiments leave the exciting possibility to tell apart the axiflavon scenario from other QCD axions. 
\newline
\newline 

\noindent{\bf  Note Added:}
During the completion of this manuscript another paper \cite{Ema:2016ops} has been submitted to the arXiv that  presents an explicit implementation of the same idea.

\vspace{0.5cm}
\noindent{\bf Acknowledgements} 
We thank T.~Gherghetta, G.~Giudice, U.~Nierste, G.~Perez, R.~Rattazzi, L.~Ubaldi and A.~Urbano for useful discussions. RZ thanks the Theory Group at DESY for kind hospitality, where this work has been partially completed. DR would like to thank CERN for the hospitality during the completion of this work.  JZ is supported in part by the U.S. National Science Foundation under CAREER Grant PHY-1151392.
\medskip

\bibliography{Axiflavon}
\end{document}